\documentclass[twocolumn,showpacs,preprintnumbers,
amsmath,amssymb]{revtex4}
\usepackage{graphicx}

\begin{document}
\title{Chemical Bond Type, Atom Packing, and Superconductivity of $\textrm{YBa}_{2}\textrm{Cu}_{3}\textrm{O}_{8-x}$
and $\textrm{Na}_{x}\textrm{WO}_{3}.$}
\author{V. N. Bogomolov}
\affiliation{A. F. Ioffe Physical \& Technical Institute,\\
Russian Academy of Science,\\
194021 St. Petersburg, Russia}
\email{V.Bogomolov@mail.ioffe.rssi.ru}
\date{\today}
\begin{abstract}
The $\textrm{YBa}_{2}\textrm{Cu}_{3}\textrm{O}_{8-x}$
superconductors are treated not as an ionic compounds but rather
as an arrays of atoms featuring both covalent and metallic
bonding. We accept as particle radii the positions of the
principal maxima in the radial distribution functions of the
charge density, which are very close to the known covalent and
metallic radii of atoms while differing strongly from the
traditional ionic radii. The graphic pattern of atomic packing in
the lattices ( approximate charge density map) reveals a number of
features inherent in nonstoichiometric compounds. The network of
atoms Y and Ba in medium Cu-O and of atoms Na in medium
$\textrm{WO}_{3}$ may be considered not as a dopant but rather as
the second component of the nanocomposite. The scheme of the onset
of the superconducting state and possible methods of its
verification are discussed.
\end{abstract}
\maketitle
\bigskip
   \section{Structures}
   One of the reasons accounting for the difficulties involved in description
   of the mechanism of superconductivity of $\textrm{YBa}_{2}\textrm{Cu}_{3}\textrm{O}_{8-x}$ and of $\textrm{Na}_{x}\textrm{WO}_{3}$ may be
   an inadequate determination of the type of chemical bonding in these compounds.
   This situation ensues actually from the traditional use of ionic radii,
   whose definition was based on a number of empirical considerations and
   hypotheses. The radius of the $\textrm{O}^{2-}$ ion is assumed equal to $1.4$\!\!\AA \;\cite{bib1,bib2,bib3},
   which automatically requires the use of ionic radii for the other atoms
   in $\textrm{YBa}_{2}\textrm{Cu}_{3}\textrm{O}_{8-x}$ and in $\textrm{Na}_{x}\textrm{WO}_{3}$ and determines the interaction type in the
   lattices. The Table lists such traditional ionic radii $r_{i}$ \cite{bib1,bib2,bib3}. Also
   presented are the radii $r_{iq}$ and $r_{aq},$ which were calculated for free ions
   and atoms as the positions of the principal maxima in the radial distribution
   functions of the charge density \cite{bib4}, together with experimental values of
   the radii of atoms in covalent compounds and metals $(r_{c}$ and $r_{m})$ \cite{bib1,bib2,bib3}.

\begin{table}
\tablename
\begin{ruledtabular}
\begin{tabular}{|l|llllll|}

\textbf{  ion}   &W$^{6+}$   &Na$^{1+}$       &Cu$^{2+}$   &Ba$^{2+}$  &Y$^{3+}$  &O$^{2-}$\\
\hline
 $r_{i}$\!\AA \cite{bib1,bib2,bib3}&0.60 &1.02 &0.73 &1.36 &0.90 &1.4\\
 $r_{iq}$\!\AA (calculated)\cite{bib4}&0.57 &0.28&0.32&0.87 &0.64&0.47\\
\hline
          \textbf{  atom}       &W               &Na &Cu  &Ba &Y &O\\
\hline
          $r_{aq}$\!\AA (calculated) \cite{bib4} &1.36 &1.72  &1.19 &2.06 &1.69 &0.45\\
          $r_{c}$\!\AA (covalent) \cite{bib1,bib2,bib3}&1.30 &1.54 &1.17 &1.98 &1.68&0.60\\
          $r_{m}$\!\AA (in metal) \cite{bib1,bib2,bib3} &1.37 &1.86 &1.28 &2.17 &1.78 &---\\
 \end{tabular}
\end{ruledtabular}
\end{table}

    As seen from the Table, the calculated atomic radii $r_{aq}$ \cite{bib4} are very close
to the covalent, $r_{c},$ and metallic, $r_{m}$ (coordination
number 6), radii derived from experimental data
\cite{bib1,bib2,bib3}. They are also in agreement with the minimum
values of the Wigner--Seitz radii (the positions where the wave
function gradient vanishes \cite{bib5}). The radius  $r_{iq} =
0.47$\!\!\AA \; of the $\textrm{O}^{2-}$ ion, however, not only
differs strongly from the generally accepted value  $r_{i} =
1.4$\!\!\AA \; but turns out to be close to the radius of the free
atom,  $r_{aq} = 0.45$\!\!\AA,\; and to the covalent radius $r_{c}
= 0.6$\!\!\AA,\; which is one half of the interatomic distance in
the $\textrm{O}_{2}$ molecule. The paramagnetism of the
$\textrm{O}_{2}$ molecule is a convincing argument for each of the
resonance quantum states,  $2p^{2} - 2p^{6} \; ({O}^{2+}-
{O}^{2-}),$ corresponding to a real physical property. In the
chalcogen series $(\textrm{n}p^{4})$ O, S, Se, Te, oxygen is the
most favorably disposed towards formation of molecular structures.

   The other ionic radii likewise differ so strongly in magnitude
   that the lattice structure with ionic radii $r_{iq}$ cannot be stable.
   At the same time, use of the calculated (free), metallic, or covalent
   radii describes substantially better not only the structures of
   $\textrm{YBa}_{2}\textrm{Cu}_{3}\textrm{O}_{8-x}$ and $\textrm{Na}_{x}\textrm{WO}_{3}$ but a number of other
   perovskite lattices too, which requires a revision of atomic interactions
   as well. For instance, in the first case\\
   $2(1r_{aq\textrm{Y}} + 2r_{aq\textrm{Ba}})  \mbox\AA = 2(1.69 + 2 \times
   2.06){\mbox\AA} = 11.61 \mbox\AA  \sim c$ $c=11.68 \mbox\AA$(\textit{c} is the cell parameter
   of
   $\textrm{YBa}_{2}\textrm{Cu}_{3}\textrm{O}_{7}$ \cite{bib6}.)
   In the $\textrm{MeB}_{6}$ compound, the $a$ lattice parameter does not change with
   a change of the metal as long as  $a = 3.68 \mbox\AA > 2r_{aqMe}$  but starts to
   increase when  $2r_{aq\textrm{Me}}$ exceeds \textit{a} \cite{bib7}. Note that bonding in the $\textrm{B}_{6}$
   octahedra remains covalent. The tungsten bronze $\textrm{Na}_{0.05}\textrm{WO}_{3}$ switches
   to superconducting state at $T_{c} = 91\textrm{K}$\cite{bib8}.  In this case,
    $a = 3.78 \mbox\AA > 2r_{aq\textrm{Na}}.$   If, however, Na is replaced by Rb or Cs,
   for which  $a < 2r_{aq \textrm{Rb}}$  or  $2r_{aq \textrm{Cs}},$  $T_{c}$ drops to a few K \cite{bib8}.
   No replacement of Na with Li or Ag  $(r_{aq\textrm{Na}}  > r_{aq\textrm{Li,Ag}})$
   was attempted. The network of atoms Na in medium $\textrm{WO}_{3}$
   may be considered not as a dopant but rather as the second
   component of the nanocomposite. Figures 1a and 1b shows
   the cell of $\textrm{WO}_{3}$ constructed using for W the atomic radius  $r_{aq\textrm{W}}.$
   For the atomic radius of oxygen we took $\sim 0.55 \mbox\AA$ which
   is close to the covalent value. Only for this value of the
   oxygen radius can the perovskite cavity accommodate freely a
   Na atom to make possible the Jahn--Teller effect (formation
   of $\textrm{Na}_{2}$ molecules with paired electrons); there will be no
   such effect, however, if the Rb and Cs atoms are compressed.
   Figures 1c and 1d present sections of the $\textrm {NaWO}_{3}$ cell along
   the $\vec{A}$ and $\vec{B}$ directions.

   Considering chemical bonding to result from interactions of
   the ionic, metallic, and covalent types, one has to accept
   in this case predominance of metallic bonding in
   the "filler"---atoms of the metal, which only weakly
   interact with the $\textrm{WO}_{3}$ "matrix" (insulator). This conclusion
   correlates well both with geometric considerations and with
   the concept of nonstoichiometric compounds to which $\textrm {Na-WO}_{3}$ belongs.

   The pattern for $\textrm{YBa}_{2}\textrm{Cu}_{3}\textrm{O}_{8}$ will be as consistent geometrically
   if one invokes the atomic rather than ionic radii for the metal
   atoms and the radius $~0.55 \mbox\AA$ for the oxygen atoms.
   Figures 2a and 2b displays a cell of the Cu--O "matrix",
   which in this case (without the "filler", i.e., the Y and Ba atoms),
   in contrast to that of $\textrm{WO}_{3},$ is certainly unstable. The atomic
   positions and the cell parameters
   of $\textrm{YBa}_{2}\textrm{Cu}_{3}\textrm{O}_{7} \;(a = 3.88 \mbox\AA,$ \;$ b = 3.83 \mbox\AA,$
   and  $c = 11.68\mbox\AA$)
   were taken from \cite{bib6}.

   \section{Superconductivity}

   A characteristic feature of a number of HTSC compounds is the
   onset of superconductivity when univalent atoms (inherently not
   superconductors), e.g., Na \cite{bib8} or Ag \cite{bib9,bib10} are inserted into
   nonmetallic matrices. These systems are typically unstable.
   Figure 3 illustrates schematically the situation with $\textrm{Na}_{0.05}\textrm{WO}_{3}.$
   For  $a > 2r_{aq\textrm{Na}}, \: \textrm{Na}_{2}$ molecules form in the lattice through
   the Jahn--Teller effect, which is impossible for the Rb
   and Cs atoms. A transition to the divalent system has occurred
   ("two-center" electron pairs appeared), with no constraints
   on superconductivity (as in the case with univalent atoms which
   do not have paired electrons) present any longer. As seen from
   Figure 2c, a similar effect is observed for the Y component
   in $\textrm{YBa}_{2}\textrm{Cu}_{3}\textrm{O}_{8}.$ In this case, both metallic components
   $(\textrm{Y}_{2}$ and Ba)
   may be considered as consisting of particles with electron pairs.
   Chains of such particles form a $3D$ network in the Cu--O matrix
   (Figure 2c, 2d). A dilute gas of diatomic molecules, for instance,
   of $\textrm{Na}_{2}$ in $\textrm{Na}_{0.05}\textrm{WO}_{3}$ (Figure 3), and, conceivably, in pressure-metallized
   condensates of inert gases (IG), for example, of $\textrm{Xe}_{2}$ in xenon, forms
   a disordered "cobweb" of superconducting filaments in an insulating
   matrix \cite{bib11,bib12}. But while the $\textrm{Na}_{0.05}\textrm{WO}_{3}$ superconductor is unstable
   against formation of domains of conventional tungsten bronze $\textrm{NaWO}_{3},$
   in xenon the $\textrm{Xe}_{2}$ molecules form an equilibrium gas.

    Consider the formation of a condensate at  $T = 0$   as particles with
    electron pairs approach one another ever closer (i.e., as their
    concentration $N$ increases). These may be, for instance, "one-center"
    electron pairs in Hg atoms or "two-center" pairs in $\textrm{Na}_{2}$ molecules.
    Because at equilibrium the condensate is a superconductor, it also
    contains electron pairs, but they exist here as "many-center"
    bosons, i.e., Cooper pairs. These "composite" bosons could conceivably
    form of atomic or molecular electron pairs whose binding energy decreased,
    and the size and overlap increased continuously as \textit{k}, the effective
    dielectric constant of the medium, grew with increasing $N.$
    In the Hg metal (the ionization potential \: $IP = 10 \textrm{eV}),$   the
    decrease of the pair binding energy is estimated to be  $(IP/kTc) \sim 10^{4}.$
    In the HTSC materials, in which the concentration of particles with
    paired electrons is lower than that in metals as a result of dilution
    by a second component, the "matrix", the effective dielectric constant
    is smaller, \: and  $T_{c},$ higher. This constant is still smaller in
    IGs.\:
    In other words,  $T_{c}$  may depend on concentration but not the bond energy and have an optimum in $N,$
    and decay of electron pairs and formation of metal
    may be preceded by Bose condensation \cite{bib12}.

    Following the process of Bose condensation of such electron pairs
    with variation of \textit{N} in many-component chemical compounds is anything
    but a simple problem. This could be possibly realized, however,
    under pressure-induced metallization of "one-component" IGs,
    which are assumed to contain diatomic excited (excimer) molecules,
    for instance, $\textrm{Xe}_{2}$ (atomic diameter  $4.6 \mbox\AA)$ immersed in a medium
    of Xe atoms in the ground state  ($5p^{6},$   diameter  $1.2 \mbox\AA$ \cite{bib4}) (atom-molecule mixture)
     \cite{bib11, bib12, bib13, bib14}). Both the pair excited states and atoms in ground state ("quantum cavities and
     channels") in IGs are as real as the paramagnetism of the $\textrm{O}_{2}$ molecules. They are
     counterparts of quantum states of atoms  $(\textrm{n}p^{5}\textrm{(n + 1)}s^{1}$ and $\textrm{n}p^{6})$
     (coexistence or superposition of two condensates based on own quantum state).
     Such an atom-molecule mixtures, perhaps, exist in $^{4}\textrm{He}$ and $^{3}\textrm{He}$.
     The dielectric constants of the IGs are on the order of a few units, so
     that the electron pairs should be bound stronger than those in an HTSC.
     By now, xenon has been reliably established to undergo metallization \cite{bib15}.
     Only its optical parameters have, however, been studied, while the
     magnetic characteristics (superconductivity) did not attract interest.
     Superconductivity was observed to set in under pressure in sulfur
     (${T_{c}} = 17\textrm{K}$ \cite{bib16}) and in oxygen $({T_{c}} = 0.6 \textrm{K}$ \cite{bib17}).
     Both sulfur and oxygen are, however, paramagnets, so that strong local magnetic
     fields could suppress the superconductivity.

    The validity of the considerations bearing on the IGs (condensates
    of particles with electron pairs in a medium of ground-state atoms)
    may be argued for by some anomalous properties of palladium.
    Palladium (Pd: ground state (g.s.)  $4d^{10}5s^{0}, \; 2r_{aq} = 1.2$\!\!\AA,  excited
    state (e.s.)  $4d^{9}5s^{1})$ in the form of a condensate  $(2r_{m}= 2.75$\!\!\AA)
    correlates with a strongly metallized IG \cite{bib18}. Unlike IGs, Pd has a
    low excitation energy  $(\sim 1\textrm{eV}).$ However, the Pd lattice still
    has some atoms in the ground state ("quantum cavities and channels"),
    which becomes manifest, for instance, in the anomalously high
    permeability of palladium for hydrogen (see, also \mbox{Nb
   (g.s.: $4d^{4}5s^{1}$ - e.s.:  $4d^{5}5s^{0}, \quad 2r_{aq}=1.50 \mbox\AA)$ and}   \mbox{Ta
    (g.s.: $5d^{3}6s^{2}$ - e.s.:  $5d^{5}6s^{0}; \quad 2r_{aq}=1.56 \mbox\AA).$}\\ As the Pd
    lattice is extended (\textit{N} decreases) through incorporation
    of foreign atoms (Li, Ag, Cu, Au, B, C, N, H etc.), Pd
    becomes superconducting with  $T_{c}$ of up to $17\textrm{K}$  \cite{bib19,bib20}.
    The same effect take place for compounds Nb-C,  Nb-Sn,  Nb-Ge   $( T_{c} \sim 11\textrm{K} \div 23 \textrm{K}).$

    The insulator--superconductor--metal transition driven by variation of
    the concentration of particles with electron pairs can be studied both
    on one-component systems (compression of an IG or "extension" of Pd)
    and, naturally, on HTSC materials. The latter are usually nonstoichiometric
    compounds, so that the type of chemical bonding prevailing in their
    components and between them depends critically on the states in which
    the atoms reside \cite{bib21}.

   Thus, the major technological difficulty in the way of developing HTSC
   materials consists in preparing stable systems of particles with paired
   electrons (for instance, divalent atoms or molecules) in concentrations
   at which the electron pairs already are capable of tunneling (Bose condensation)
   while not yet separating into single electrons, either as a result of conventional
    chemical interactions or through a decrease in their binding energy in an effective
    dielectric medium. Such an intermediate state corresponds
    to a "stretched" or "diluted" substance, which can be stabilized
    by using "solid solvents" as matrices. In some cases such nonequilibrium
    concentrations are possibly "frozen" by chance fluctuations in the
    technology of synthesis \cite{bib8,bib9,bib10}. Low and equilibrium concentration of
    particles with two paired electrons may exist in some compounds
    (like
    $\mathrm {Ag_{2}[Ag_{3}Pb_{2}O_{5}]})$
    with 1D channels \cite{bib10}.  In searching for efficient methods
    of structural stabilization (with help of some amount of inert
    particles as "additional diluent" (third component), for instance ),
    a better understanding of the interactions in the structure of each
    component is of crucial importance. Such methods of stabilization
    can be found only by using realistic atomic and ionic radii rather
    than conventional quantities, which are nothing else but a consequence
    of postulated bonding types. The matrices in such nanocomposites also
    exist in an unusual state because of the unusual structure and contact
    interaction with the second component. This complicates greatly the
    investigation and description of such systems.

\begin{figure*}[tbp]
\includegraphics[width=\linewidth]{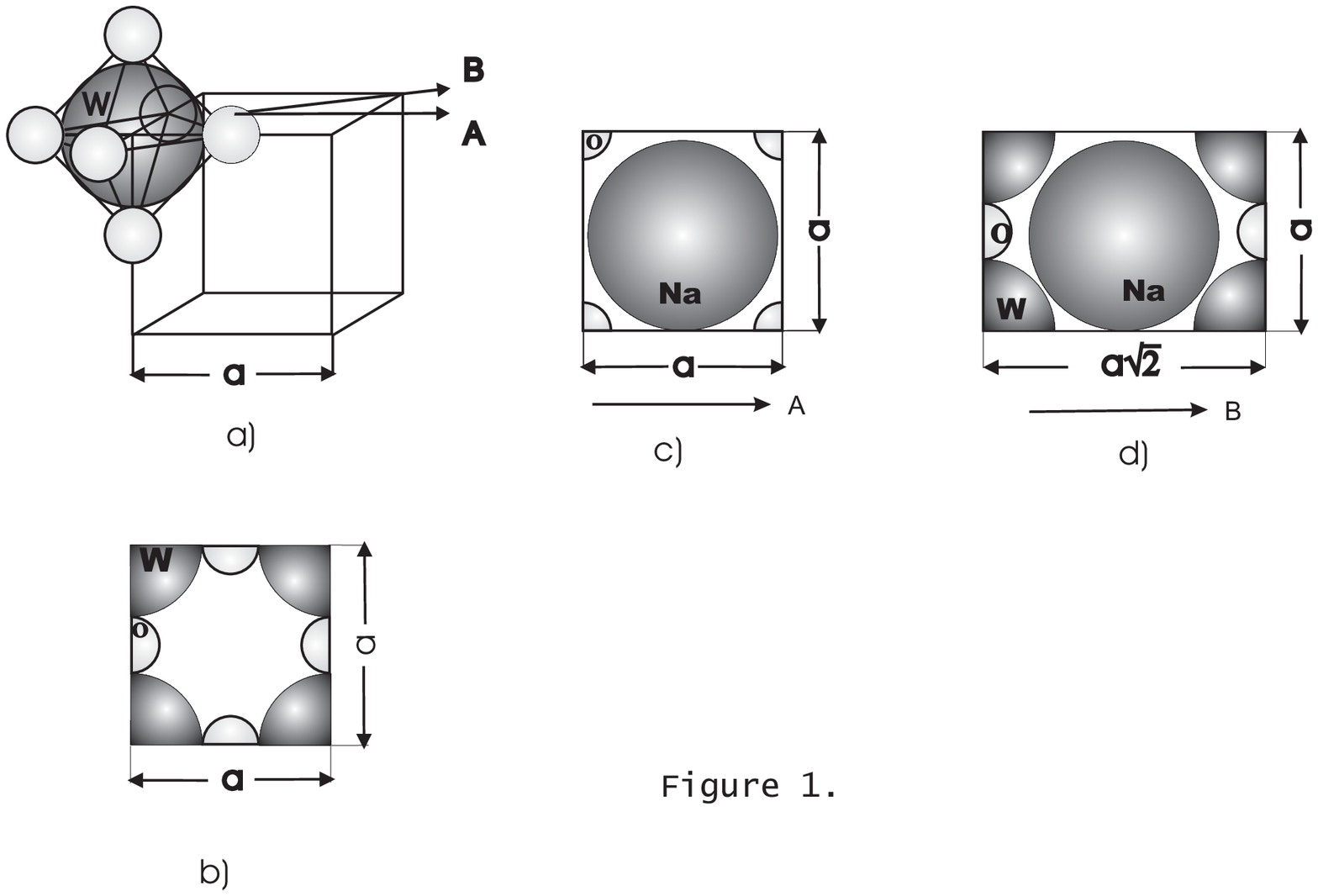}
\caption{\\(a) $\textrm{WO}_{3}$ cell; \\(b) Side face of the
$\textrm{WO}_{3}$ cell;\\ (c) Section of $\textrm{NaWO}_{3}$ cell
along the $\vec{A}$ direction;
\\(d) Section of $\textrm{NaWO}_{3}$ cell along the
$\vec{B}$ direction.} \label{fig1}
\end{figure*}

\begin{figure*}[tbp]
\includegraphics[width=\linewidth]{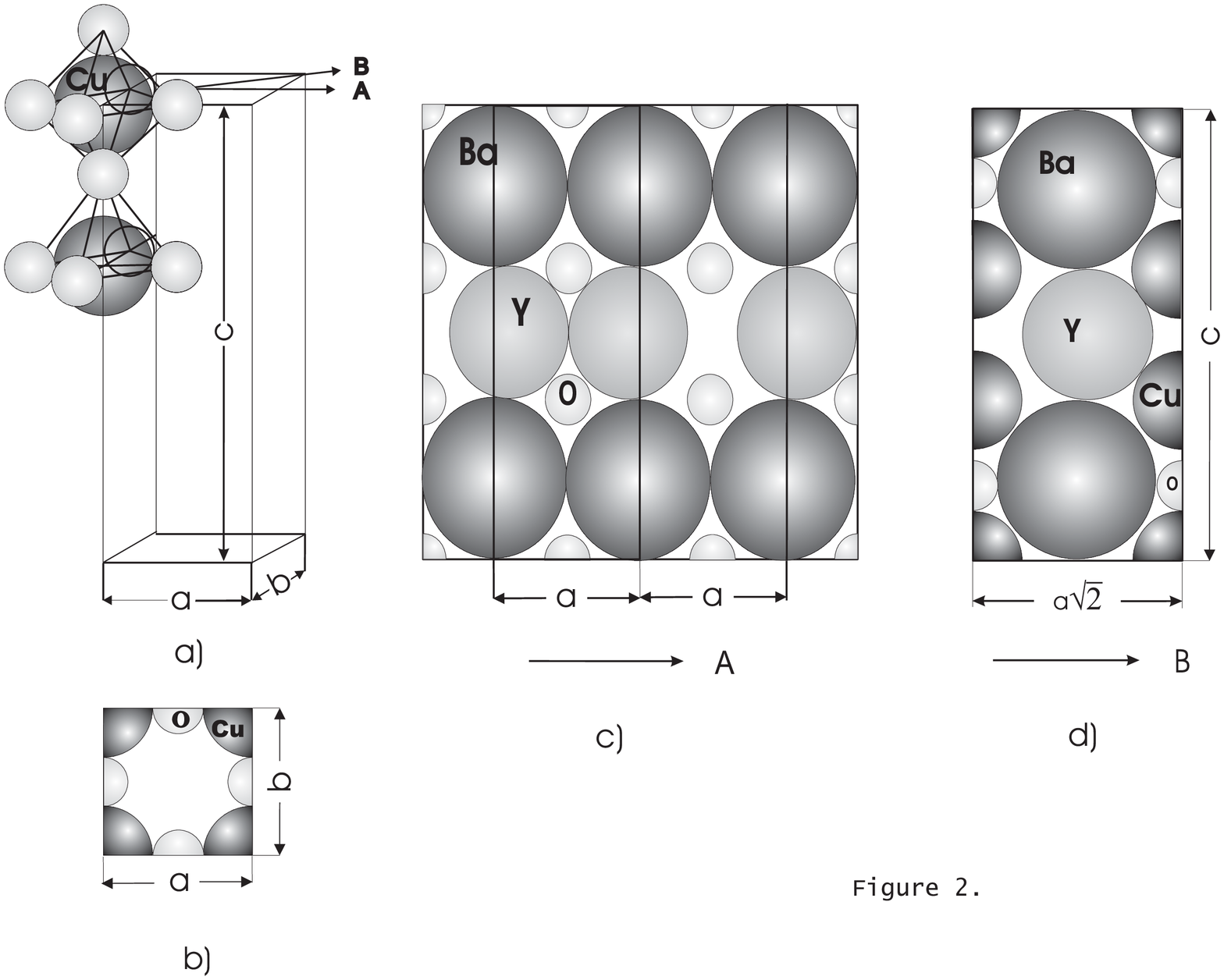}
\caption {\\(a). $\textrm{YBa}_{2}\textrm{Cu}_{3}\textrm{O}_{8}$
cell;
\\(b) Side face of the
$\textrm{YBa}_{2}\textrm{Cu}_{3}\textrm{O}_{8}$ cell; \\(c)
Section of $\textrm{YBa}_{2}\textrm{Cu}_{3}\textrm{O}_{8}$ cell
along the $\vec{A}$ direction; \\(d) Section of
$\textrm{YBa}_{2}\textrm{Cu}_{3}\textrm{O}_{8}$ cell along the
$\vec{B}$ direction.}  \label{fig2}
\end{figure*}

\begin{figure*}[tbp]
\includegraphics[width=0.5\linewidth]{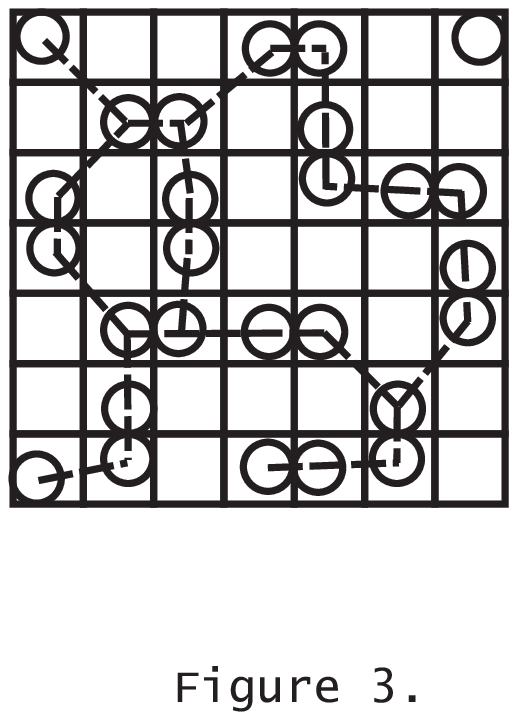}
\caption{\\Schematic arrangement of
$\textrm{Na}_{2}(\textrm{Xe}_{2})$ molecules in
$\textrm{Na}_{0.05}\textrm{WO}_{3}$ (Xe under pressure). A variant
of divalent impurity in insulator. Dashed lines: "cobweb"
\cite{bib11,bib22} of superconducting filaments of
$\textrm{Na}_{2}(\textrm{Xe}_{2})$ molecules. The system
$\textrm{Na}_{0.05}\textrm{WO}_{3}$ is unstable against formation
of metallic domains of the tungsten bronze
$\textrm{Na}_{1}\textrm{WO}_{3}.$ A "gossamer" of superconducting
filaments in insulator supposed to be in cuprates
(Fig.2)\cite{bib23}.} \label{fig3}
\end{figure*}

\end{document}